\documentclass{PoS}
\title{Development and operations of INFN optical modules for the SCT Telescope camera proposed for the Cherenkov Telescope Array Observatory}
\usepackage{todonotes}
\usepackage{amssymb}
\usepackage{comment}

\ShortTitle{INFN optical modules for the SCT Telescope camera proposed for the CTA Observatory}

\author{
C.~Adams~$ˆ{a}$,
G.~Ambrosi~$ˆ{b}$, 
M.~Ambrosio~$ˆ{c}$,
C.~Aramo~$ˆ{c}$,
W.~Benbow~$ˆ{d}$,
B.~Bertucci~$ˆ{be}$,
E.~Bissaldi~$ˆ{fg}$,
M.~Bitossi~$ˆ{h}$,
A.~Boiano~$ˆ{c}$,
C.~Bonavolont\`a~$ˆ{c}$,
R.~Bose~$ˆ{i}$,
A.~Brill~$ˆ{a}$,
J.~H.~Buckley~$ˆ{i}$,
M.~Caprai~$ˆ{b}$,
C.~E.~Covault~$ˆ{j}$,
L.~Di~Venere~$ˆ{g}$,
Q.~Feng~$ˆ{k}$,
E.~Fiandrini~$ˆ{be}$,
A.~Gent~$ˆ{l}$,
N.~Giglietto~$ˆ{fg}$,
F.~Giordano~$ˆ{fg}$,
R.~Halliday~$ˆ{j}$,
O.~Hervet~$ˆ{m}$,
G.~Hughes~$ˆ{d}$,
T.~B.~Humensky~$ˆ{a}$,
M.~Ionica~$ˆ{b}$,
W.~Jin~$ˆ{n}$,
P.~Kaaret~$ˆ{o}$,
D.~Kieda~$ˆ{p}$,
B.~Kim~$ˆ{q}$,
F.~Licciulli~$ˆ{g}$,
S.~Loporchio~$ˆ{fg}$,
V.~Masone~$ˆ{c}$,
T.~Meures~$ˆ{r}$,
B.~A.~W.~Mode~$ˆ{r}$,
R.~Mukherjee~$ˆ{k}$,
A.~Okumura~$ˆ{s}$,
N.~Otte~$ˆ{l}$,
F.~R.~Pantaleo~$ˆ{fg}$,
R.~Paoletti~${ht}$,
A.~Petrashyk~$ˆ{a}$,
J.~Powell~$ˆ{n}$,
K.~Powell~$ˆ{l}$,
D.~Ribeiro~$ˆ{a}$,
A.~Rugliancich~$ˆ{h}$,
M.~Santander~$ˆ{n}$,
R.~Shang~$ˆ{q}$,
B.~Stevenson~$ˆ{q}$,
L.~Stiaccini~$ˆ{ht}$,
L.~P.~Taylor~$ˆ{r}$,
\speaker{L.~Tosti}~$ˆ{be}$,
V.~Vagelli~$ˆ{bev}$,
M.~Valentino~$ˆ{uc}$,
J.~Vandenbroucke~$ˆ{r}$,
V.~Vassiliev~$ˆ{q}$,
P.~Wilcox~$ˆ{o}$,
D.~A.~Williams~$ˆ{m}$
for the CTA Consortium \footnote{for collaboration list see PoS(ICRC2019)1177}
\newline \\
\llap{$ˆa$} Physics Department, Columbia University, New York, NY 10027, USA\\
\llap{$ˆb$} INFN Sezione di Perugia, Perugia, Italy\\
\llap{$ˆc$} INFN Napoli, Italy\\
\llap{$^d$} Center for Astrophysics, Harvard \& Smithsonian, Cambridge, MA 02138, USA\\
\llap{$ˆe$} Universit\`a degli Studi di Perugia, Perugia, Italy\\
\llap{$ˆf$} Dipartimento Interateneo di Fisica dell'Universit\`a e del Politecnico di Bari\\
\llap{$ˆg$} INFN Bari, Via E. Orabona 4, 70125 Bari, Italy\\
\llap{$ˆh$} INFN Sezione di Pisa, Pisa, Italy\\
\llap{$ˆi$} Department of Physics, Washington University, St. Louis, MO 63130, USA\\
\llap{$ˆj$} Department of Physics, Case Western Reserve University, Cleveland, Ohio 44106\\
\llap{$ˆk$} Department of Physics and Astronomy, Barnard College, Columbia University, NY 10027, USA\\
\llap{$ˆl$} School of Physics \& Center for Relativistic Astrophysics, Georgia Institute of Technology, 837 State Street NW, Atlanta, GA 30332-0430, USA\\
\llap{$ˆm$} Santa Cruz Institute for Particle Physics and Department of Physics, University of California, Santa Cruz, CA 95064, USA\\
\llap{$ˆn$} Department of Physics and Astronomy, University of Alabama, Tuscaloosa, AL 35487, USA\\
\llap{$ˆo$} Department of Physics and Astronomy, University of Iowa, Iowa City, IA 52242, USA\\
\llap{$ˆp$} Department of Physics and Astronomy, University of Utah, Salt Lake City, UT 84112, USA\\
\llap{$ˆq$} Department of Physics and Astronomy, University of California, Los Angeles, CA, USA\\
\llap{$ˆr$} Department of Physics and Wisconsin IceCube Particle Astrophysics Center, University of Wisconsin, Madison, WI 53706, USA\\
\llap{$ˆs$} Institute for Space--Earth Environmental Research and Kobayashi--Maskawa Institute for the Origin of Particles and the Universe, Nagoya University, Nagoya 464-8601, Japan\\
\llap{$ˆt$} Dipartimento di Scienze Fisiche, della Terra e dell'Ambiente, Universit\`a degli Studi di Siena, Siena, Italy\\
\llap{$ˆu$} CNR-ISASI, Italy\\
\llap{$ˆv$} Now at ASI Italian Space Agency - Scientific Research Unit, Roma, 00133, Italy\\

E-mail: \email{luca.tosti@pg.infn.it}}

\abstract{
The Schwarzschild-Couder Telescope (SCT) is a proposal for the Medium Size Telescopes of the Cherenkov Telescope Array. Its concept is based on a two-mirror optical system designed to improve the telescope field of view and image resolution with respect to the single mirror Davies-Cotton solution. The SCT camera is planned to be instrumented with 177 photodetection modules, each composed of 64 Silicon Photomultiplier (SiPM) pixels. The third generation of $6\times 6~mm^2$ high density NUV SiPMs (NUV-HD3) produced by Fondazione Bruno Kessler (FBK) in collaboration with INFN has been used to
equip optical units to be integrated on the upgrade of the camera of the SCT prototype (pSCT). Each optical unit is composed of an array of 16 NUV-HD3 SiPMs coupled with the front-end electronics, which is designed for full-waveform nanosecond readout and digitization using the TARGET-7 ASIC. Several optical units have been assembled and tested in the laboratories of INFN and have been integrated on the camera of the pSCT telescope, that is currently operating at the Fred Lawrence Whipple Observatory. In this contribution we report on the development, assembly and calibration of the optical units that are currently taking data on the pSCT camera.
}

\FullConference{36th International Cosmic Ray Conference -ICRC2019-\\
		July 24th - August 1st, 2019\\
		Madison, WI, U.S.A.}
\begin{document}
\section{The Schwarzschild-Couder Medium Size Telescope (SCT) proposal}
Cherenkov Telescope Array (CTA) will be the largest ground-based gamma-ray observatory for very high energy gamma rays. CTA will be installed in two sites, one in the northern hemisphere (La Palma in Spain) and one in the southern hemisphere (Atacama desert in Chile). It will be equipped with more than $100$ telescopes in three different configurations: Large, Medium and Small-Sized Telescopes (LST, MST and SST) and it will be fully operational by $2025$.
A novel design of a Schwarzschild-Couder dual mirror optics for the MST telescopes has been proposed \cite{sct} for the CTA baseline array. The dual mirror solution improves the image resolution with respect to the single mirror Davies-Cotton (DC) implementation, and the more focused optics allows to equip the camera with a dense array of Silicon Photomultipliers (SiPMs) as image pixels. An MST-SCT array  ultimately results in an improved angular resolution and off-axis sensitivity for the CTA array when compared with a conventional MST-DC array. 
A pathfinder prototype telescope based on this design, the pSCT telescope, has been successfully constructed at the Fred Lawrence Whipple Observatory (its inauguration was on $17$th January $2019$) \,\cite{vassilev_psct_icrc2019}. The operations of SCT prototype (pSCT) are providing important information to develop and establish the procedures for the optical alignment \cite{feng_psct_icrc2019} and for the calibration of the SiPM camera\,\cite{taylor_icrc_2019} to confirm the possibility of an extended SCT array for CTA.

\begin{figure}
\centering
     \includegraphics[width=.3\textwidth]{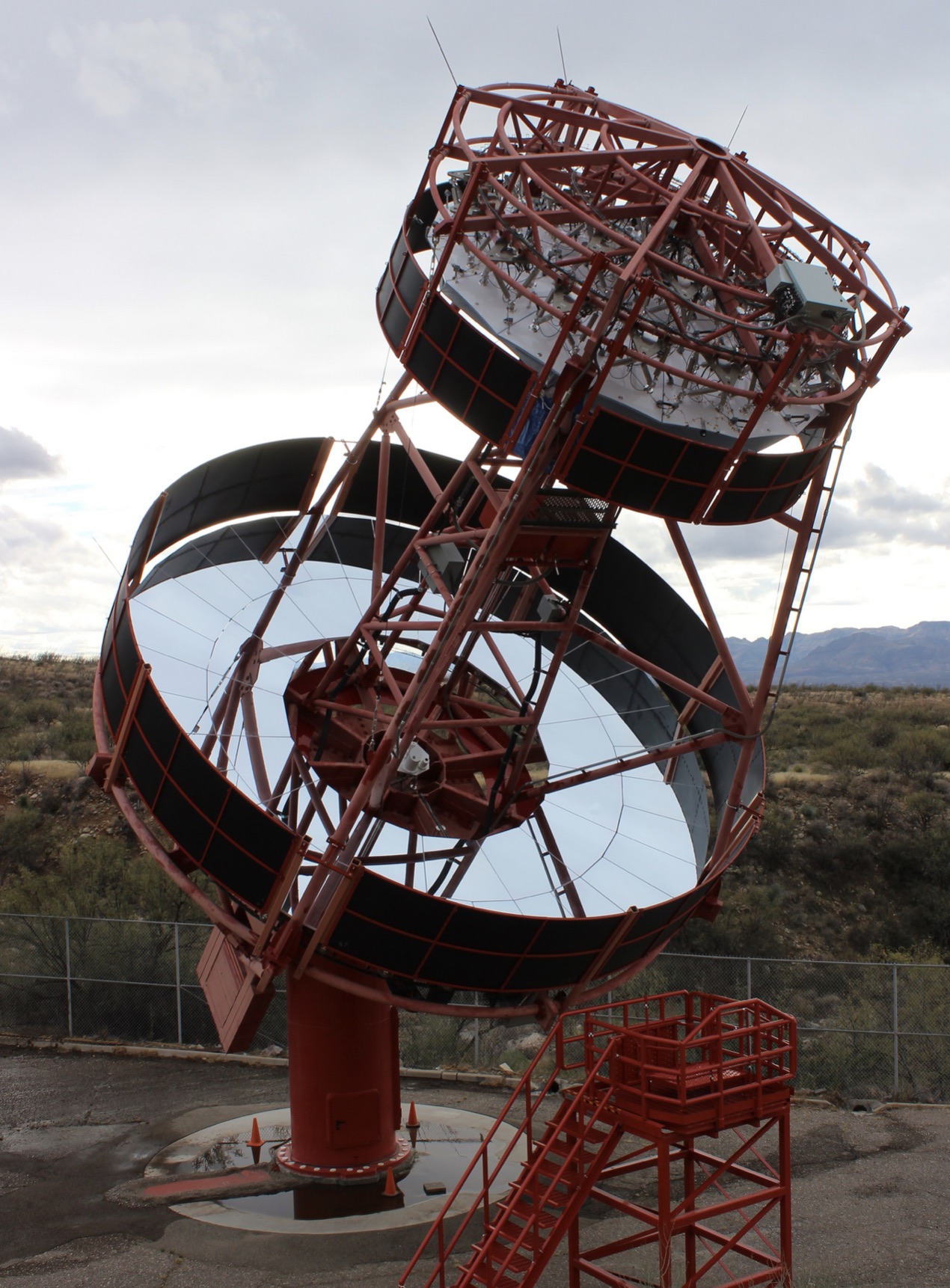}
     \includegraphics[width=.6\textwidth]{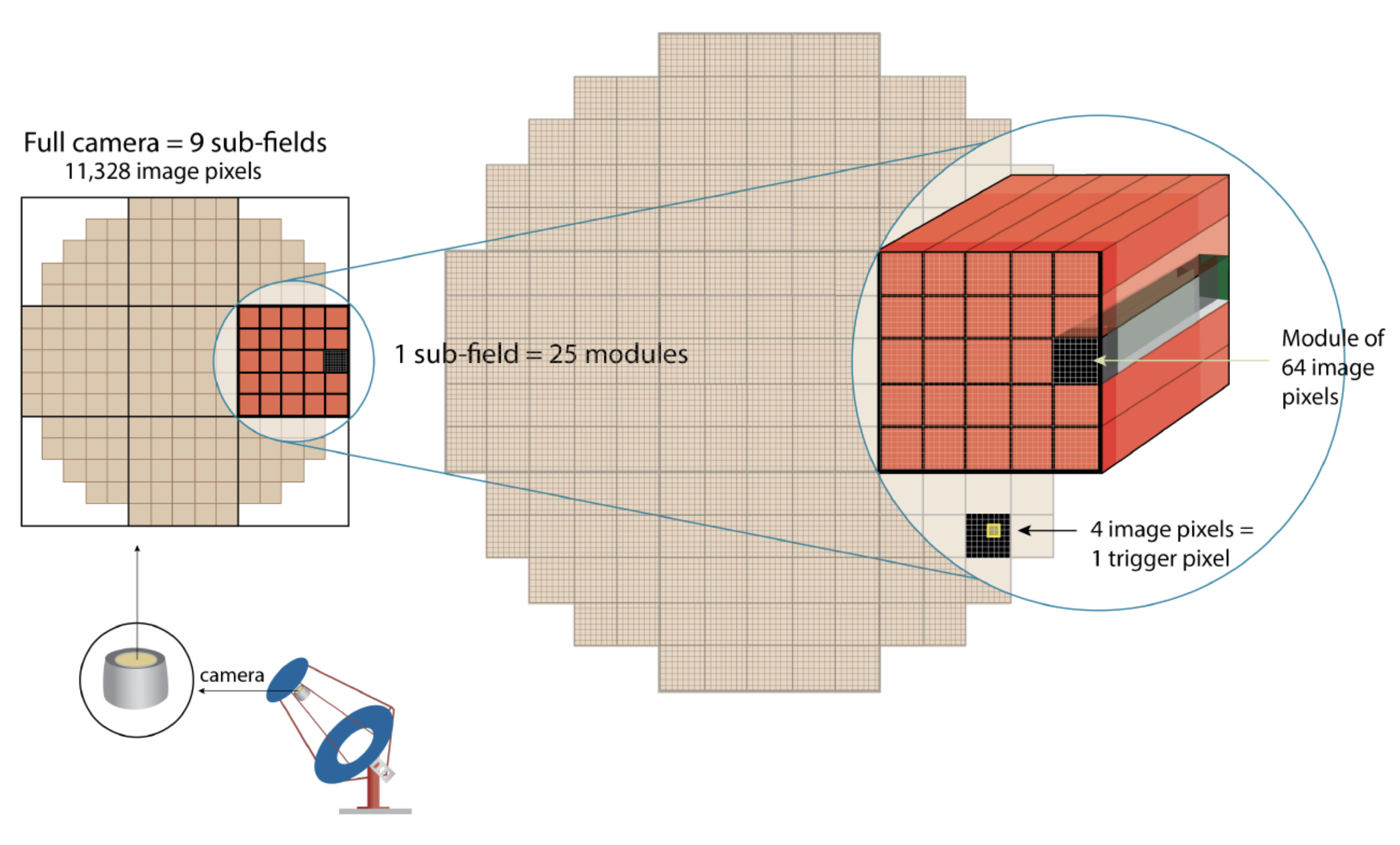}
     \caption{Left: The pSCT telescope fully integrated in the Fred Lawrence Whipple Observatory. Right: Scheme of the pSCT camera.}
     \label{SCT}
\end{figure}

\subsection{The pSCT telescope camera}

\begin{figure}
\centering
     \includegraphics[width=.6\textwidth]{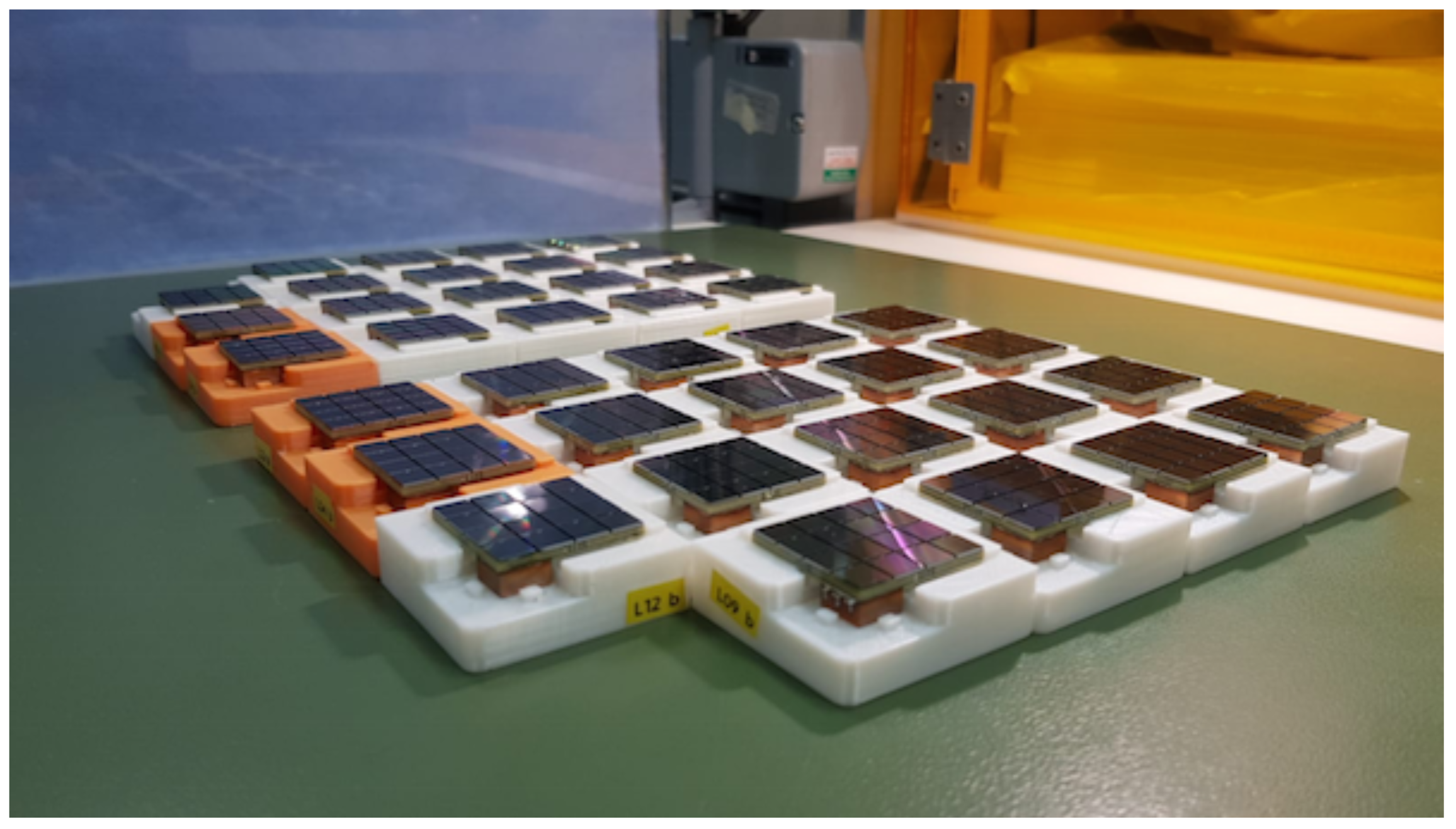}
     \caption{batch of optical units based on NUV-HD3 SiPMs ready for the integration on the pSCT camera}
     \label{single_matrix}
\end{figure}

The improved point spread function (PSF) of the SCT telescope allows the sky image over a field of view of approximately $8$ degrees to be focused over a compact camera equipped with SiPMs. The pSCT camera has a width of 81\,cm over an area of $0.43\, m^2$ equipped with $11328$ SiPMs, each representing an image pixel corresponding to a $0.067$ degree in the sky. 
The entire focal plane camera is arranged in $9$ sub-fields each with up to $25$ optical modules, for a total of $177$ optical modules; $4$ of the $9$ sub-field are partially filled to produce the circular shape required for the imaging. 
Each module is composed of $4$ basic optical units with $16$ pixels each. Fig.\ref{SCT} shows, on the right, a conceptual scheme of the camera. 
The pSCT camera readout concept is based on a TARGET-7 ASIC front-end electronics. The TARGET-7 ASIC handles the first level trigger - based on the signal of $4$ adjacent sensors -  and digitization for $16$ channels, sampling the analog signal at $1GS/s$ with a switched capacitor array of $16384$ units for a maximum buffer depth of $\sim\,16\,\mu s$. The backplane of the camera handles and manages the data acquisition for separate camera sectors. The pSCT camera is currently equipped with 16 optical modules based on Hamamatsu S12642 MPPCs. In order to investigate a possible upgrade of the pSCT camera\,\cite{meures_psct_icrc2019}, the remaining $9$ modules of the same sector have been instead equipped with the third generation of Near Ultra Violet High Density SiPMs (NUV-HD3) produced by Fondazione Bruno Kessler (FBK) in collaboration with Istituto Nazionale di Fisica Nucleare (INFN) (See Fig.\ref{single_matrix}). 
In this contribution we describe the performances of FBK NUV-HD3 SiPMs, their integration and calibration on the pSCT camera to establish the possibility to equip the full pSCT camera with upgraded electronics and sensors based on this technology.
\section{Performances of FBK NUV-HD3 Silicon Photomultipliers}
\begin{figure}
\centering
     \includegraphics[width=.5\textwidth]{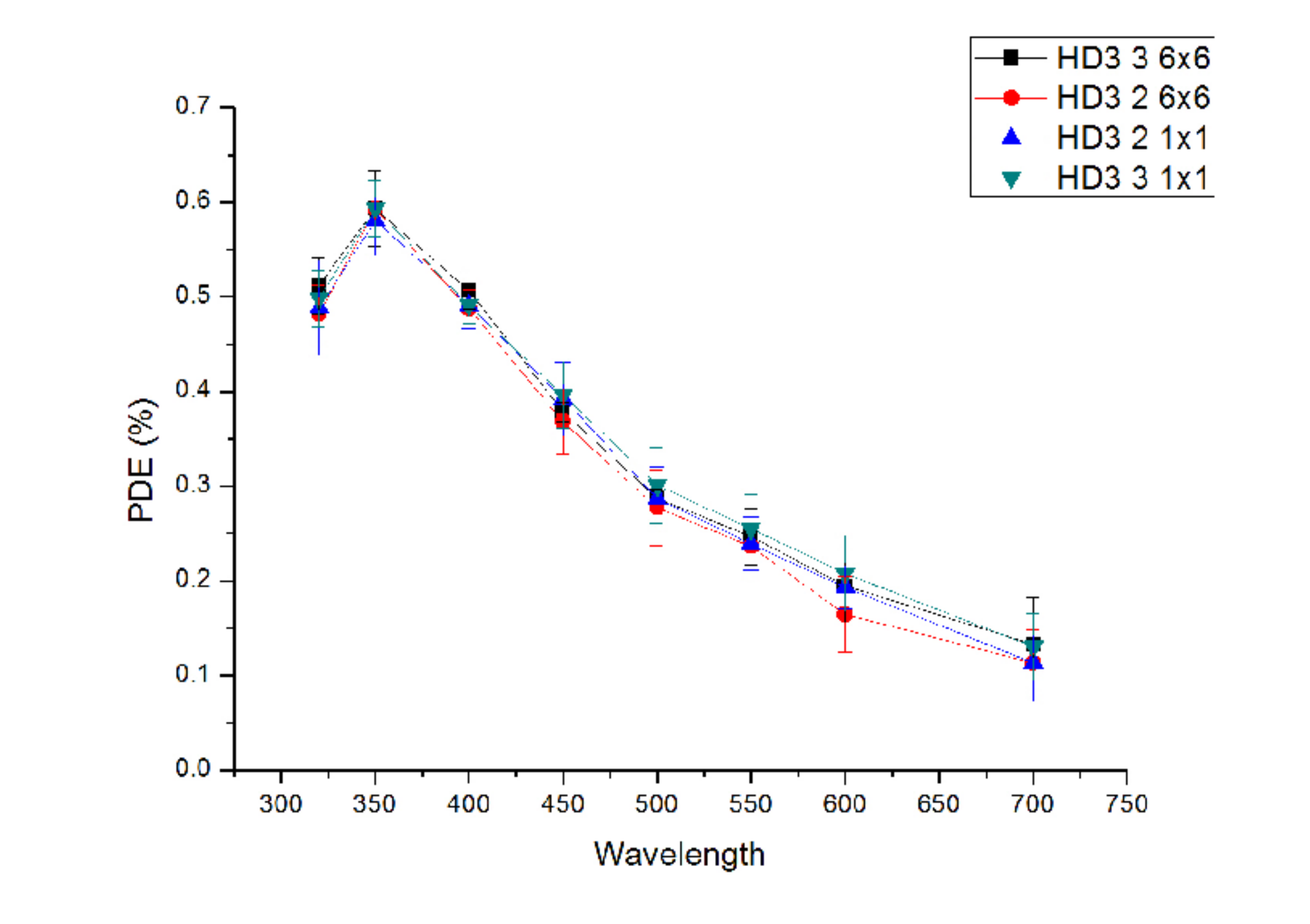}
     \includegraphics[width=.4\textwidth]{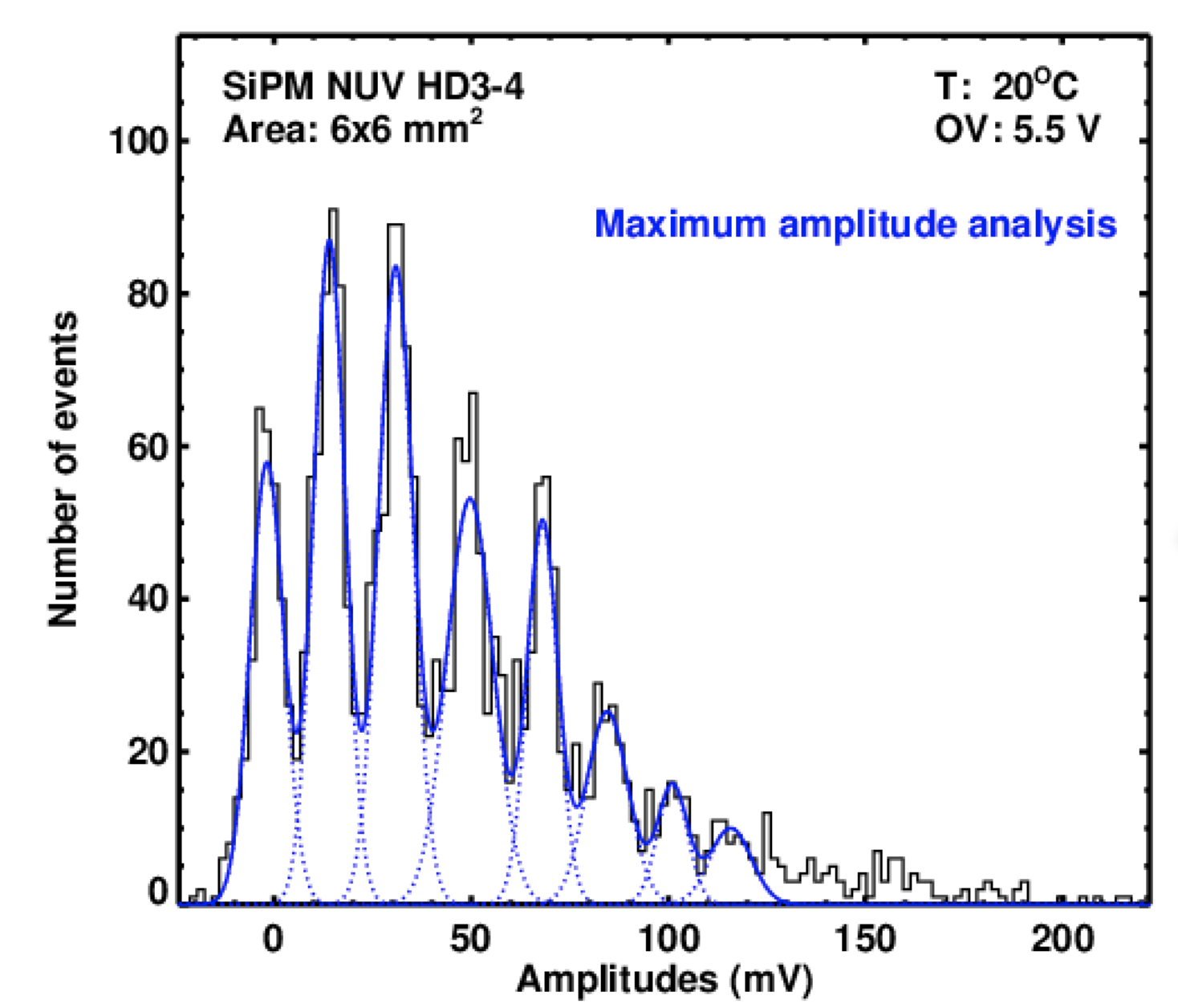}
     \caption{Left: Wavelength dependence of the PDE (maximum about 350$nm$) for different type and dimensions (in $mm$) of NUV-HD3 . Right: Distribution of the signal amplitude (for a single SiPM) in standard condition (T$= 20^{\circ} C$; OV = $5.5 V $)}
     \label{characterization}
\end{figure}

NUV-HD3 SiPMs have been developed by FBK in collaboration with INFN for optimized applications in CTA telescopes. They are based on the FBK Low-CT SiPM design, described in details in\,\cite{FBK_Sensors_2019}, optmizing their sensitivity for the detection of Cherenkov light while minimizing the amount of optical crosstalk to improve the telescope effective area at low energies. Several productions of NUV-HD SiPMs have been tested in the laboratories of INFN, and the third production has been selected for integration in the pSCT camera.\\
The active area of NUV-HD3 SiPMs amounts to $6\times6~mm^{2}$, with a microcell area of $40\,\times\,40\,\mu m^2$. They feature a breakdown voltage of approximately $27\,V$ at room temperature with a temperature gradient of $\sim 30\,{mV}/^{\circ}C$, a Photo Detection Efficiency (PDE) peak of $\sim 50\% $ at $350\,nm$ dropping below $20\%$ above $500\,nm$ and a single photon dark count rate below $150\,kHz/mm^2$ at $20^{\circ} C$. The details of the laboratory measurements and procedures are discussed in details elsewhere \cite{misure}. An example of the results of the characterization of the SiPMs PDE and signal amplitude are shown in Fig.\ref{characterization}.
\section{Assembly and characterization of SiPM arrays for pSCT}
\begin{figure}[ht]
\centering
     \includegraphics[width=.8\textwidth]{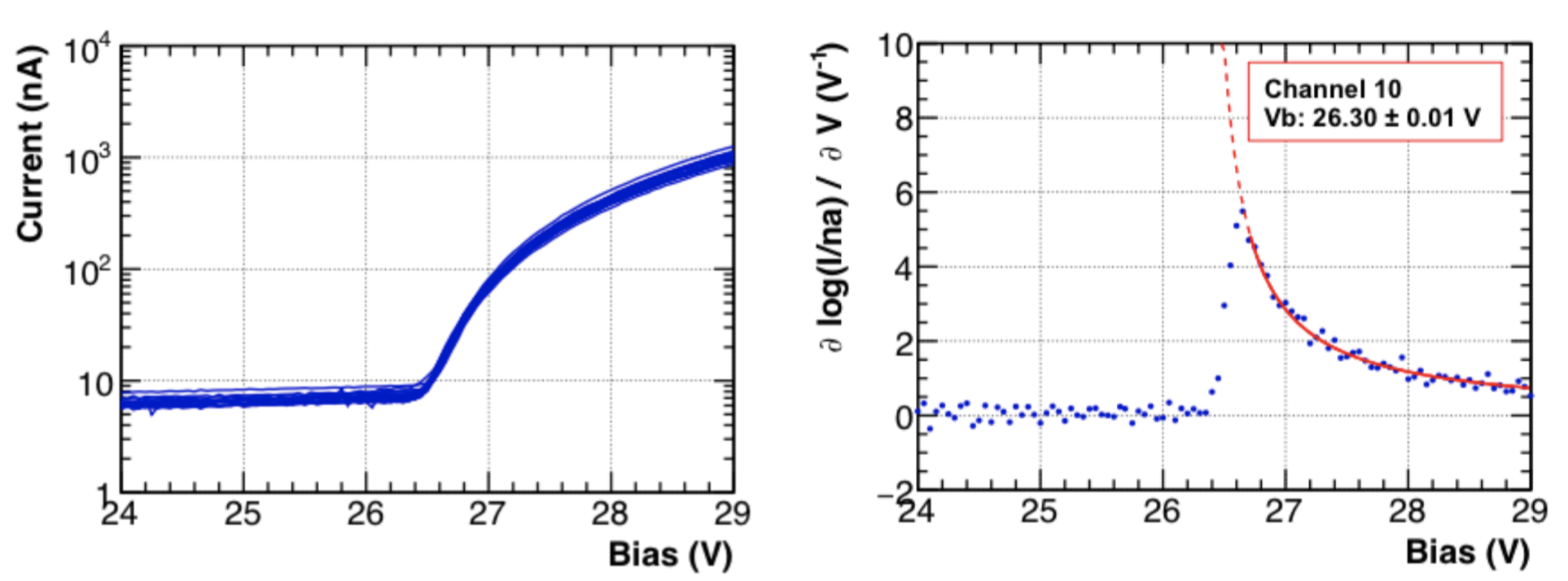}
     \caption{Left: Voltage-current (IV) curves for the 16 channels in an optical unit. Right: fit to the derivative of the logarithm of the current for one channel for the estimation of the breakdown voltage $V_b$.}
     \label{iv}
\end{figure}

\begin{figure}
\centering
     \includegraphics[width=.52\textwidth]{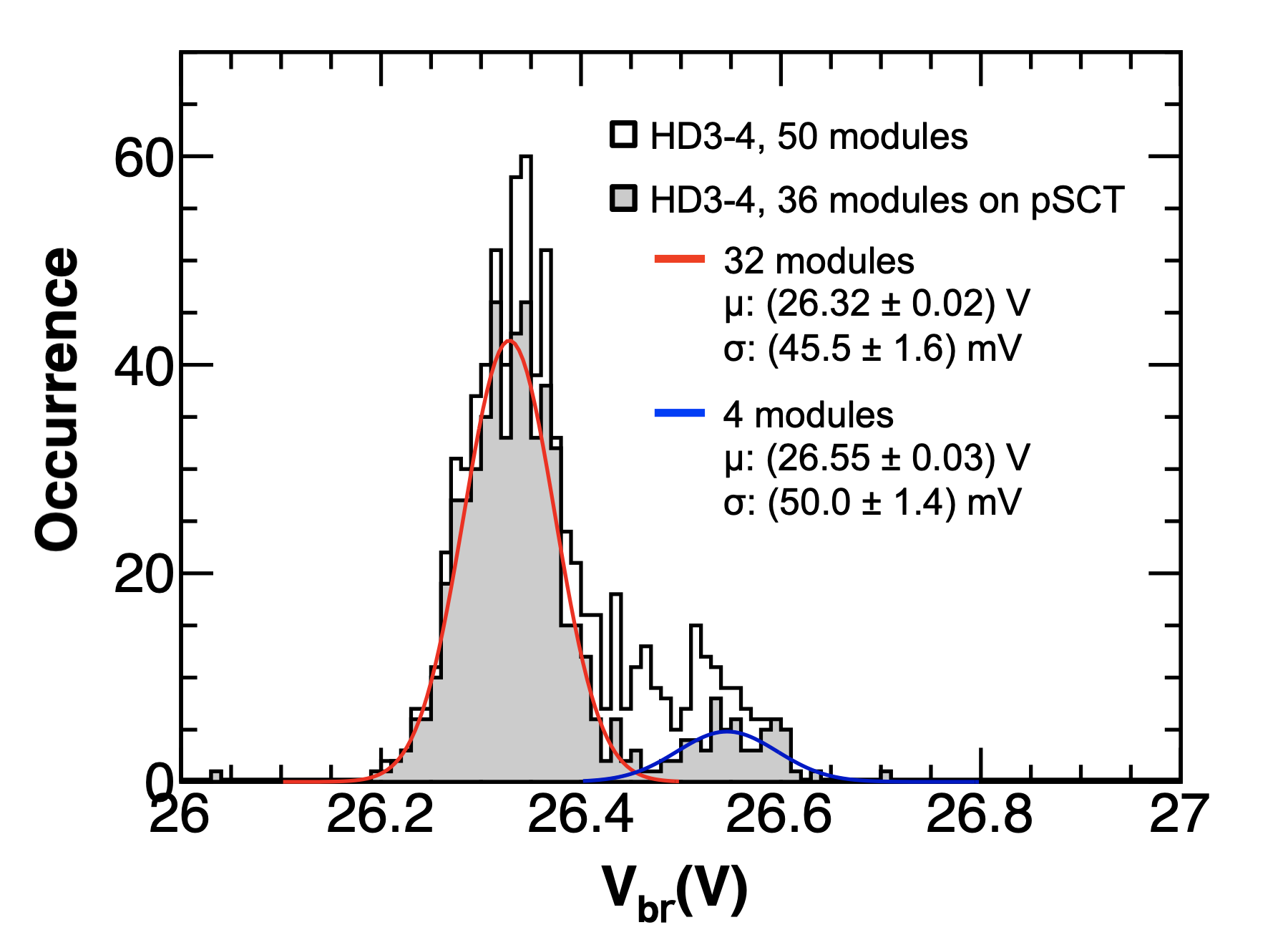}
     \caption{Distribution of the breakdown voltage value of the entire batch of $800$. The red and blue distributions represent the composition of the SiPM substrates.}
     \label{distribution}
\end{figure}

While single sensors are provided by the vendor, the procedures for the assembly and packaging of the optical modules have been completely developed by INFN.\\Each 64 SiPM module covers an area of $54\,\times\,54\,mm^{2}$ and is divided in $4$ optical units each composed of $16$ pixels. INFN has designed custom $27\,\times\,27\,mm^{2}$ area PCBs with $0.5\,mm$ sensor-sensor distance to obtain uniform pixel coverage of the modules and of the camera, and compatible with the pSCT camera design. A dedicated assembly procedure has been developed by INFN 
to integrate the sensors on the photodetection units\,\cite{ricap_2018}.\\
After assembly, the quality of the SiPM sensors on each module has been tested via the measurement of the characteristic voltage-current (IV) curve, that measures the current drain of the sensor as function of the inverse bias voltage. The value of $V_{b}$ for each SiPM is estimated with a fit on the derivative of the logarithm of the current I(V) with a function of the bias voltage as shown in Eq. \ref{ILD}
\begin{equation}\label{ILD}
\frac {\partial{log~(I(V))}}{\partial V} = \frac{2}{V - V_{b}}
\end{equation}
An example of IV curves for $16$ sensors of a module and of the procedure to measure the breakdown voltage value is shown in Figure\,\ref{iv}.\\
The procedure has been repeated for the optical units intended to be integrated on the pSCT telescope.
The distributions of the measured breakdown voltage for $800$ NUV-HD3 SiPM channels is shown in Figure\,\ref{distribution}. Out of $50$ units, $36$ have been chosen to equip the pSCT telescope camera. The distribution for their breakdown voltages is represented by the grey histogram. Depending on the silicon tape of implantation, the average values of the breakdown voltage differ by approximately $200\,mV$.
After the assembly, the performances of all the modules have been tested with a dedicated electronics setup\,\cite{readoutchain}. The $16$ SiPMs of each module have been illuminated with a $380\,nm$ pulsed laser. The charge signal of the $16$ sensors has been acquired using a CAEN V792 QDC over a fixed integration time of $50\,ns$. The output signal of each SiPM has been consequently amplified using a custom 16-channel front end board to match the dynamic range of the QDC.
\begin{figure}
\centering
     \includegraphics[width=.48\textwidth]{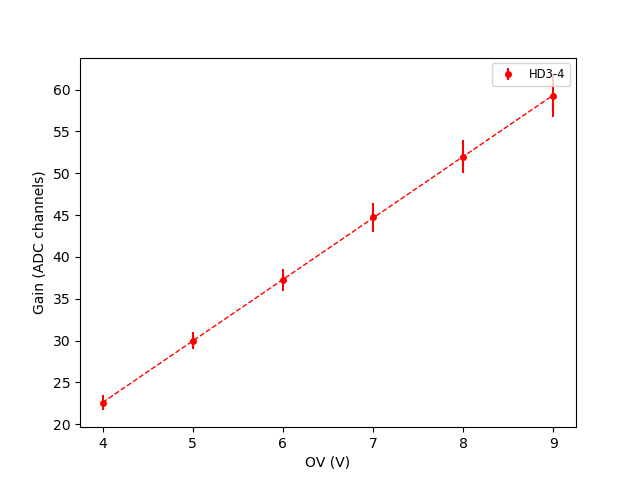}
     \includegraphics[width=.48\textwidth]{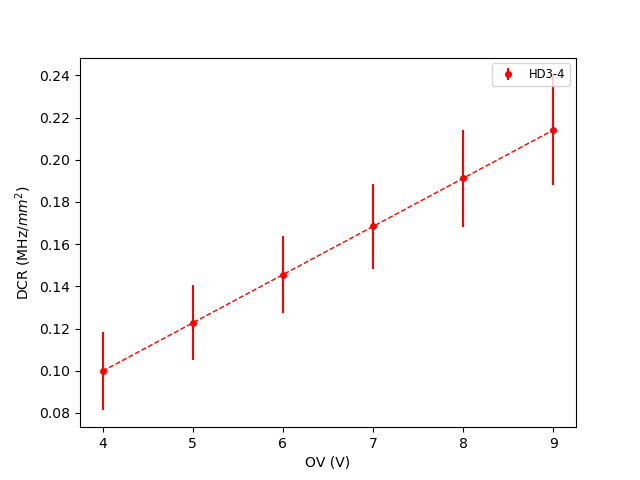}
     \caption{Gain and DCR performances (as function of the overvoltage applied to the sensor) of the optical units. Each point represent the mean value and the std. deviation of the distribution of all tested modules}
     \label{dcr}
\end{figure}
 The S/N ratio, dark count rate and gain (Fig. \ref{dcr}) are tested and, according to the results and the $V_b$ estimation, the optical units are selected to equip the camera of the SCT prototype (pSCT). This kind of test are performed in the INFN laboratories of Bari.
The optical units are clustered in groups of $4$ and then coupled with the TARGET-7 ASIC readout and trigger modules. This electronics was developed to allow the sampling and  waveform digitization of $16$ independent channels (for a signal time of $16 \mu s$) at $1~GSample/s$. Moreover the TARGET-7 ASIC can perform a first-level trigger decision based on the signal of 4 adjacent pixels. The operation of the TARGET-7 internal trigger has been verified by varying the intensity of the light for a fixed trigger threshold (amplitude scans) and by varying the trigger threshold for a fixed light intensity (threshold scans) \cite{trigger}. Pedestal calibration, waveform acquisition and the trigger efficiency tests were performed in Bari and Pisa.
\section{Installation on the camera}
\begin{figure}[!h]
\centering
     \includegraphics[width=.8\textwidth]{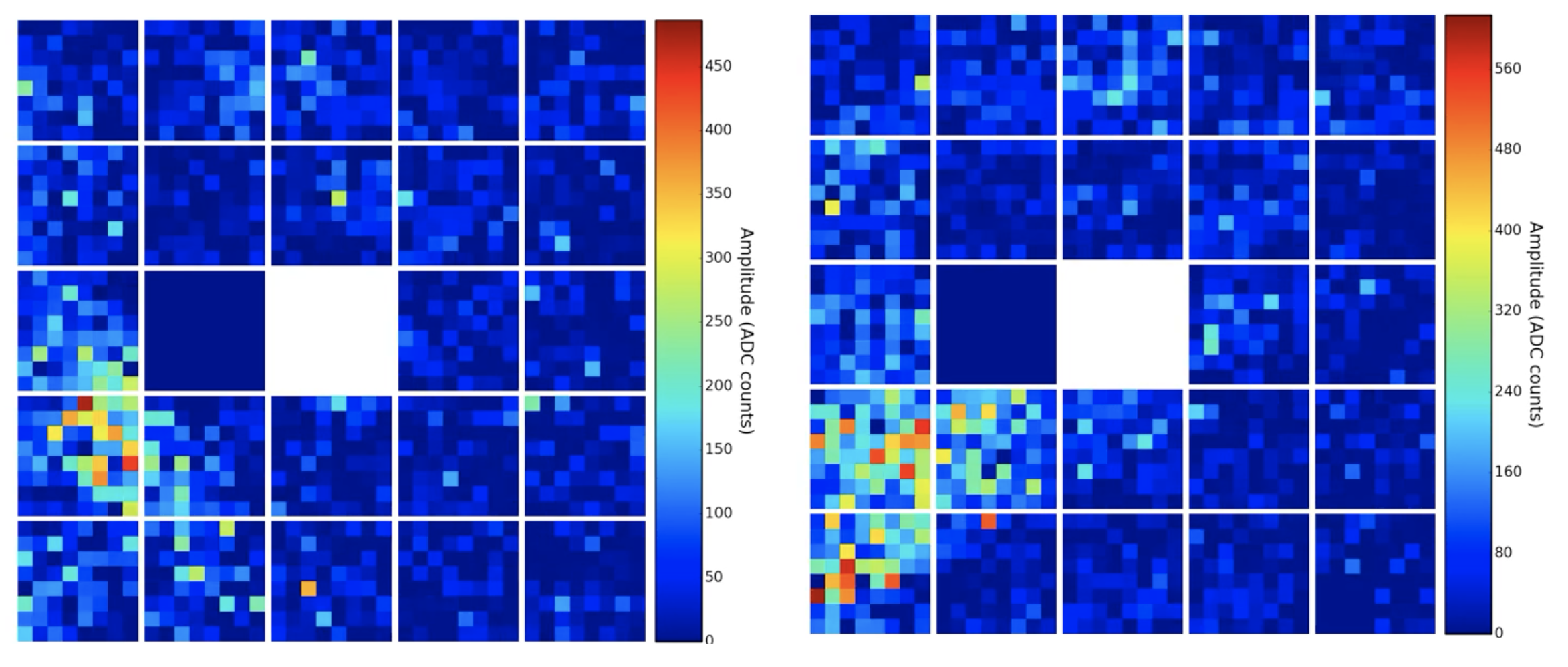}
     \caption{
     Examples of first lights detected on $23$rd of January $2019$}
     \label{first_lights}
\end{figure}

The optical units (clustered by four) have been aligned to be integrated with the camera using a reference frame with a tolerance of $0.02 mm$ and specials copper elements. This approach allows the curvature of the camera sectors to be set and improve part of the aberration correction. These procedures have been performed at the Georgia Institute of Technology (Atlanta, USA). The aligned units (coupled with the TARGET-7 modules) have been integrated on the pSCT telescope camera frame at University of Wisconsin (Madison, USA) or directly on site at Fred Lawrence Whipple Observatory (AZ, USA). A total of $150$ optical units (prototypes included) have been assembled and tested in the INFN laboratories and, as discussed before, $50$ of them have the right level of performances for the integration on the telescope camera; $36$ of them have been selected to build $9$ complete modules for SCT, for a total of $576$ SiPMs. The units have been already installed in the pSCT camera together with 16 modules equipped with the main technology based on Hamamatsu MPPCs. The purpose is to test the performances of FBK NUV-HD3 SiPM in standard telescope operations to verify the opportunity to equip a fraction of the SCT telescopes for CTA with this kind of sensors. In January $2019$, the pSCT prototype was successfully inaugurated in Fred Lawrence Whipple Observatory (USA) and in the $23$rd of January $2019$ first lights were be detected. Now the pSCT prototype is currently taking data for commissioning and calibration of the camera (Fig. \ref{first_lights}).
\newpage
\section{Major prospect}
In $2019$, one additional sector of the camera ($25$ modules, $1600$ SiPMs) is planned to be integrated with FBK NUV-HD3 SiPMs. An additional technological advance for this modules is represented by the development of new ASIC boards based on the TARGET design. The new concept separates the sampling and digitization on a TARGET C chip and the trigger functions on the companion T5TEA ASIC to reduce the trigger noise \cite{target7}. An additional preamplifier ASIC (the SMART chip) is intended to further decouple the analog preamplification from the digitization process. This board is currently being prototyped for the pSCT by INFN. The new Front End Electronics (FEE) design is planned to be integrated with the pSCT backplanes together with the complete integration of the $177$ optical modules of the pSCT camera in the near future.

\section*{Acknowledgements}
This work was conducted in the context of the CTA pSCT Project. We gratefully acknowledge the financial support from the agencies and organizations listed in
\begin{footnotesize}
https://www.cta-observatory.org/consortium\_acknowledgments/
\end{footnotesize} 
 and from Progetto Premiale TECHE.it

\end{document}